# Schools of skyrmions with electrically tunable elastic interactions

Hayley R. O. Sohn[1], Changda D. Liu[1], and Ivan I. Smalyukh[1,2,3*]

*[1]Department of Physics and Materials Science and Engineering Program, University of Colorado, Boulder, CO 80309, USA*

*[2]Department of Electrical, Computer, and Energy Engineering and Soft Materials Research Center, University of Colorado, Boulder, CO 80309, USA*

*[3]Renewable and Sustainable Energy Institute, National Renewable Energy Laboratory and University of Colorado, Boulder, CO 80309, USA*

*\* Correspondence to: <u>ivan.smalyukh@colorado.edu</u>*

**Coexistence of order and fluidity in soft matter often mimics that in biology, allowing for complex dynamics and applications like displays. In active soft matter, emergent order can arise because of such dynamics. Powered by local energy conversion, this behavior resembles motions in living systems, like schooling of fish. Similar dynamics at cellular levels drive biological processes and generate macroscopic work. Inanimate particles capable of such emergent behavior could power nanomachines, but most active systems have biological origins. Here we show that thousands-to-millions of topological solitons, dubbed "skyrmions", while each converting macroscopically-supplied electric energy, exhibit collective motions along spontaneously-chosen directions uncorrelated with the direction of electric field. Within these "schools" of skyrmions, we uncover polar ordering, reconfigurable multi-skyrmion clustering and large-scale cohesion mediated by out-of-equilibrium elastic interactions. Remarkably, this behavior arises under conditions similar to those in liquid crystal displays and may enable dynamic materials with strong emergent electro-optic responses.**

**Introduction.**

Soft matter and living systems are commonly described as close cousins[1], both with properties stemming from interactions between the constituent building blocks that are comparable in strength to thermal fluctuations. Active soft matter systems[2,3] are additionally out-of-equilibrium in nature, like everything alive. They exhibit emergent collective dynamics that closely mimic such behavior in living systems[2-4]. For example, mechanically-agitated fluidized monolayers of rods form a dynamic granular liquid crystal (LC)[5]. Coherent motion emerges in many systems where particles communicate through collisions or short-range interactions like screened electrostatic repulsions[6-22]. However, these interactions typically cannot be tuned in strength or switched from attractive to repulsive. Moreover, with a few exceptions[5-7,10,11,19-22], including the ones in which electric energy is used to power motions[6,7,11], most active matter systems have biological origins and either chemical or mechanical energy conversion within the constituent building blocks. This poses the grand challenge to develop versatile reconfigurable active matter formed by inanimate, man-made particles both as models of biological systems and for technological uses[21].

We describe an emergent collective dynamic behavior of skyrmions[23-25], particle-like two-dimensional (2D) topological analogs of Skyrme solitons used to model atomic nuclei with different baryon numbers[26,27]. In LCs[1,28], these skyrmions are elements of the second homotopy group[29] and contain smooth but topologically-nontrivial, spatially-localized structures in the alignment field of constituent rod-like molecules, the director field $\mathbf{n}(\mathbf{r})$. They are characterized by integer-valued topological invariants, the skyrmion numbers[29]. Depending on the applied voltage, the internal $\mathbf{n}(\mathbf{r})$-structures within our skyrmions adopt different orientations relative to the 2D sample plane and the far-field alignment. Thousands-to-millions of skyrmions start from random orientations and motions while each individually converting energy due to oscillating voltage, but then synchronize motions and develop polar ordering within seconds. The ensuing schools of topological solitons differ from all previously studied systems[2-22] because skyrmions have no physical boundaries, membranes, chemical composition or density gradients, or singularities in the order parameter at the level of the host fluid[30-36], even though they exhibit giant number fluctuations in terms of the skyrmionic, topologically protected, localized structures of $\mathbf{n(r)}$. Although the LC medium is nonpolar, spontaneous symmetry breaking and many-body dynamic interactions lead to polar ordering of skyrmions, which is characterized by near-unity values of polar and velocity order parameters. Electrically tunable interactions stemming from the orientational elasticity of LCs[30,31] provide a versatile means of controlling this behavior while probing order and giant number fluctuations within the schools. The dynamic multi-skyrmion assemblies echo formation of high-baryon-number skyrmions in nuclear physics due to addition of charge-one topological invariants, which in equilibrium condensed matter could be only achieved when forming skyrmion bags[29], very differently from the behavior of singular active matter topological defects that conserve the net winding number to always add to zero[2,33-36]. Our findings highlight the interplay between nonsingular topology of field configurations and out-of-equilibrium behavior and promise a host of technological uses.

**Results.**

**Skyrmion schooling.** Schooling of fish (Fig. 1a), like many other forms of collective motions[2], is accompanied by inhomogeneities and dynamic local clustering. Similar behavior is observed in our rather unusual schools formed by thousands-to-millions of localized particle-like skyrmionic orientational structures of $\mathbf{n(r)}$ within LCs (Fig. 1b-f). While moving and bypassing obstacles, these skyrmions exhibit dynamically self-reconfigurable assembly (Fig. 1f and Supplementary Movie 1). In our experiments, under conditions and sample preparation similar to that in LC displays, such skyrmions are controllably mass-produced at different initial densities (Methods)[32] and also generated one-by-one using laser tweezers[23,29]. Skyrmion stability is enhanced by soft perpendicular boundary conditions on the inner surfaces of confining glass plates (Fig. 1d)[23] and the LC's chirality, which prompts $\mathbf{n(r)}$ twisting[23,24]. At no fields, the structure of each skyrmion is axisymmetric (Fig. 1g)[23], with π-twist of $\mathbf{n(r)}$ from the center to periphery in all radial directions and containing all possible $\mathbf{n(r)}$-orientations within it. Due to the used LC's negative dielectric anisotropy, electric field $\mathbf{E}$ applied across the cell tends to align $\mathbf{n(r)}{\perp}\mathbf{E}$ (Fig. 1d), so that $\mathbf{n(r)}$ around the skyrmions progressively tilts away from the cell normal with increasing voltage $U$, whereas $\mathbf{n(r)}$ within the skyrmions morphs from an originally axisymmetric structure (Fig. 1g) to a highly asymmetric one (Fig. 1j,k) that matches this tilted director surrounding. Since the tilting of $\mathbf{n(r)}$ in $\mathbf{E}$ with respect to the sample plane breaks the nonpolar symmetry of the resulting effectively-2D structure, we vectorize $\mathbf{n(r)}$ and visualize it with arrows colored by orientations and corresponding points in the two-sphere $\mathbb{S}^2$ order parameter space (Fig. 1g)[29]. The asymmetric skyrmion is described by a preimage vector connecting preimages of the south and north poles of $\mathbb{S}^2$ (Fig. 1k), the regions where $\mathbf{n(r)}$ points into and out of the sample plane, respectively. The skyrmion number, a topological invariant describing how many times $\mathbf{n(r)}$ within the single skyrmion wraps around $\mathbb{S}^2$, remains equal to unity, indicating topological stability with respect to smooth $\mathbf{n(r)}$-deformations prompted by changing $U$. This is confirmed by experimental polarizing optical micrographs of individual skyrmions at different $U$ (Fig. 1i,m) that closely match their computer-simulated counterparts (Fig. 1h,l). Skyrmions exhibit only Brownian motion at no fields and at high frequencies (like 1kHz) of applied field, at which $\mathbf{n(r)}$ cannot follow the temporal changes of $\mathbf{E}$ and responds to its time average.[24] At frequencies for which the voltage oscillation period $T_U = 1/f$ is comparable or larger than the LC response time, spatial translations of individual skyrmions[24] in an oscillating electric field arise because the temporal evolution of asymmetric $\mathbf{n(r)}$ is not invariant upon reversal of time with turning $U$ on and off within each $T_U$ (Figs. 1g-m and 2a,b). Because of the initial axisymmetric structure of skyrmions and homeotropic $\mathbf{n(r)}$ background around them, the spontaneous symmetry breaking leads to random motion directions of individual skyrmions within the sample plane. This feature of our system indicates that the coherent unidirectional motion of many skyrmions within the thousands-to-million schools is an emergent phenomenon with a physical mechanism that relies on inter-skyrmion interactions, which we explore in detail below.

**Collective motions of two-to-hundred skyrmions**. To gain insights into the emergent schooling with electrically reconfigurable clusters of topological solitons (Fig. 1g-m), we first probe how the dynamics of skyrmions change with increasing number density (Fig. 2). Using laser tweezers, we set up a "race" by arranging skyrmion pairs along a straight line together with single skyrmions (Fig. 2a) and then start oscillating $U$ at $f = 2$ Hz by effectively turning it on and off every 0.5s while using a 1kHz carrier frequency electric source. The single solitons move faster than pairs (Fig. 2a) and trio-chains (Supplementary Movie 2), whereas large assemblies of 40 and more skyrmions barely move (Supplementary Fig. 1). Since spatial translations of skyrmions arise from temporal evolution of asymmetric $\mathbf{n(r)}$ not invariant upon reversal of time with turning $U$ on and off, skyrmions within large assemblies tend to share these asymmetric distortions to reduce the elastic free energy[30,31], which impedes their translations while in tightly packed reduced-asymmetry assemblies. However, with the oscillating field's modulation changing to $f = 1$ Hz, the solitons tend to spread apart, so that the motion of well-spread pairs is then nearly as fast as that of single skyrmions (Fig. 2b). The dynamic evolution of $\mathbf{n(r)}$ during motions of skyrmion pairs is enriched by elastic interactions that tend to minimize the ensuing elastic free energy costs due to arranging skyrmions at different relative positions and can be controlled from attractive to repulsive (Fig. 2c). Tuning $U$ and $f$ alters these inter-skyrmion interactions and reconfigures larger kinetic assemblies of tens (Supplementary Movie 3) and hundreds of skyrmions (Supplementary Movie 4), which in turn alters their dynamics. For example, with tuning $f$ within 2-8Hz, a cluster of 29 skyrmions shown in Supplementary Movie 3 re-forms into a long chain, which breaks into smaller clusters and then rearranges again and again, multiple times. Hundreds of skyrmions at initial packing fractions <0.01 (Fig. 2d-f) form chains meandering like snakes (Fig. 2d-f, Supplementary Fig. 2 and Movie 4) and bounce from each other within a dynamic cluster-like region. Inaccessible to skyrmions at equilibrium[29], such self-reconfigurable behavior emerges in both right- and left-handed chiral LCs obtained with different chiral additives (Methods). Moreover, over long periods of time, there is no repetition of assemblies and trajectories of the randomly-directed motion, which we verify by analyzing the complex dynamics (Fig. 2f and Supplementary Movies 3 and 4). Although the polarizing video microscopy reveals how the asymmetric periodically changing $\mathbf{n}(\mathbf{r})$ evolution powers motions of multi-skyrmion assemblies (Fig. 2),[24] this mechanism alone cannot explain self-reconfigurable randomly-directed dynamics of two-to-hundreds skyrmion assemblies and schooling of thousands-to-millions of skyrmions with tunable clustering within the schools. Systematic analysis of pair interactions (Fig. 2c) reveals that this complex behavior can be understood by taking into account the out-of-equilibrium elastic interactions between skyrmions that arise from minimizing the elastic free energy due to partial sharing of $\mathbf{n}(\mathbf{r})$-distortions associated with multiple moving skyrmions, as we detail below.

**Out-of-equilibrium elastic interactions**. Elastic interactions between skyrmions emerge to reduce the free energy costs of $\mathbf{n}(\mathbf{r})$-distortions around these topological solitons, like in nematic colloids[29-32], albeit typically without the dynamic $\mathbf{n}(\mathbf{r})$ fully reaching equilibrium because of the voltage modulation and soliton motions. The elastic interactions between skyrmions confined to a 2D plane have dipolar nature[29], though the complex temporal evolution of $\mathbf{n}(\mathbf{r})$ in periodically modulated $U$ makes these elastic dipoles effectively change their tilt relative to the 2D sample plane within $T_U$ and self-propel while they interact. Such dynamic dipolar skyrmions mutually repel at small $U$, but exhibit anisotropic interactions (Fig. 2c), including attractions, when oscillating $\mathbf{E}$ prompts their $\mathbf{n(r)}$ symmetry breaking (Fig. 1j,k) and motions. Oscillating $\mathbf{E}$ rotates preimage dipoles from pointing orthogonally to the sample plane at $U$=0 (Fig. 1g) to being tilted or in-plane (Fig. 1j,k) when $U$ increases, with the effective tilt periodically changing with a voltage modulation period (Fig. 1d) typically comparable to the LC's response time. When released at different relative initial positions using laser tweezers[32], skyrmions with parallel preimage vectors perpendicular to substrates always repel whereas skyrmions with in-plane preimage dipoles attract when placed head-to-tail and repel when side-by-side (insets of Fig. 2c). Depending on $U$, $f$ and relative skyrmion positions, the strength of reconfigurable elastic pair interactions (Fig. 2c) varies within (1-10,000)$k_\text{B}T$, where $k_\text{B}$ is the Boltzmann constant and $T$ is the absolute temperature. Since the response of $\mathbf{n(r)}$ to oscillating $U$ is fast on the timescales of skyrmion motions at ~1μm per second, tuning $\mathbf{n(r)}$ by $U$ and $f$ modifies elastic forces between parallel dipoles by changing the effective tilt (averaged over $T_\text{U}$) of the dipole moments relative to the sample plane.

**Emergence of polar order and coherent motions.** In the presence of thousands-to-millions of skyrmions (Fig. 1b-f), applied $\mathbf{E}$ initially induces random tilting of the director around individual skyrmions, so that their south-north preimage unit vectors $\mathbf{p}_i = \mathbf{P}_i/|\mathbf{P}_i|$ point in random in-plane directions (Fig. 3). Individual skyrmions exhibit translational motions with velocity vectors $\mathbf{v}_i$ roughly antiparallel to their $\mathbf{p}_i$. With time, coherent directional motions emerge (Supplementary Movies 1 and 5-7), with schooling of skyrmions either individually-dispersed (Fig. 4 and Supplementary Movie 5) or in various assemblies (Fig. 5 and Supplementary Movies 1, 6 and 7). Velocity and polar order parameters $S$=$|\sum_i^N \mathbf{v}_i|/(N\mathbf{v}_s)$ and $Q$=$|\sum_i^N \mathbf{p}_i|/N$ characterize degrees of ordering of $\mathbf{v}_i$ and $\mathbf{p}_i$ within the moving schools[19], where $N$ is the number of skyrmionic particles and $\text{v}_s$ is the absolute value of velocity of a coherently-moving school. Both $S$ and $Q$ increase from zero to ~0.9 within seconds (Fig. 4d), indicating the emergence of coherent unidirectional motion of polar skyrmionic particles, like that of fish in schools[2,3]. At relatively low initial packing fractions (~0.1 by area) we observe no clustering of skyrmions as they move coherently within the schools, repelling each other at short distances and weakly attracting at larger distances (Fig. 4c). This emergent behavior is different from pair interactions and dynamics at similar voltages (Fig. 2c), where moving skyrmions tend to attract to form chains at shorter inter-skyrmion distances. The presence of such short-range repulsive and long-range attractive interactions is consistent with the formation of coherently-moving schools and results from many-body interactions (Fig. 4c), where elastic interactions between skyrmions with periodically-evolving $\mathbf{n}(\mathbf{r})$ are further enriched by backflows and electro-kinetic effects.[28] As the elastic interactions vary from attractive to repulsive within $T_U$, the effective time-averaging of these interactions localizes skyrmionic particles at distances roughly corresponding to the distance at which pair interactions are comparable to $k_BT$ (Figs. 2c and 4a,c). The many-body interactions between skyrmions then lead to effective cohesion within the school and their coherent collective motion. Using video microscopy, we also analyze the mean $<N>$ and root mean square $\Delta N = <(N-<N>)^2>^{1/2}$ of particles within different sample areas (Methods and Fig. 4e). Unlike in the case of random Brownian motion of colloidal particles or the same skyrmions with $\alpha=0.5$, when $\Delta N \propto <N>^{1/2}$, skyrmions in schools exhibit giant number fluctuations with $\Delta N \propto <N>^{\alpha}$, where $\alpha=0.763$ (Fig. 4e), as well as fluctuations in the local number density probed by counting the numbers of skyrmions within a selected sample area versus time (Fig. 4f).

**Tunable clustering, edges and cohesion in skyrmion schools**. We alter the skyrmion schooling behavior by inducing formations of clusters (Fig. 5). At moderately large 0.1-0.4 skyrmion packing fractions within the schools (Fig. 5), we observe motions of dynamically self-assembled clusters at $U$=2.5-3.75V and linear chains at $U$=3.75-4.5V (Fig. 5a-d and Supplementary Movies 1, 6 and 7). Like individual skyrmions in schools of lower density (Fig. 4), moving clusters and chains remain separated at distances ~30μm corresponding to effective pairwise interactions ~$k_BT$ (Fig. 4c). During this schooling, small clusters and linear polar chains exhibit giant number fluctuations with varying values of α=0.61-0.85 (Fig. 5e). Within the clusters and chains, skyrmions are kept at separation distances comparable to their lateral size (Fig. 5a-m) and roughly consistent with the separation distances corresponding to minima of potentials of pair interactions at similar conditions (Fig. 2c), which can be tuned by $U$ and $f$ through tuning the temporal evolution of $\mathbf{n}(\mathbf{r})$, as we show for the case of chains in Fig. 5n. This dynamic assembly of multi-skyrmions echoes nuclear physics models, where subatomic particles with high baryon numbers can be modeled as clusters of elementary skyrmions[27]. Each skyrmion cluster can be characterized by a net skyrmion number corresponding to a sum of topological invariants of elementary skyrmions within it (e.g. clusters in Fig. 5f and j have net skyrmion numbers of 7 and 5, respectively). Tuning packing fractions, $U$ and $f$ allows for emergence of a broad range of this collective behavior (Fig. 5). The direction of collective motion within inch-square cells (Figs. 1b-d, 4 and 5) is selected spontaneously and emerges only at sufficiently large number densities of skyrmions, though gradients of cell gap thickness and external fields could potentially be used to control it.

Edges of schools are well defined regardless of the internal clustering within the schools (Fig. 6a-d). Individual skyrmions, which happen to be slowed down by imperfections (Fig. 6e), move faster than the clusters and thus "catch up" to the school's edge. Although the speed decreases as the number of skyrmions within the clusters increases, this reduction is less than 50% even for very large clusters containing over 100 skyrmions (Fig. 6f), also showing how the dynamic behavior under schooling conditions differs from that of individual skyrmions and their small-to-medium clusters (Supplementary Fig. 1).

**Diagram of dynamic and static states.** We summarize the schooling behavior of skyrmions using a structural diagram (Fig. 7), where we present results for square-wave electric fields oscillating at frequency $f$ (we note that the diagram of states changes when other electric signal waveforms and various carrier frequencies are used, although exploration of all these parameter spaces is outside the scope of our present work). Being unstable at high $U$ and low $f$ (Fig. 7), skyrmions exhibit static self-assemblies at low $U$ and high $f$ and dynamic structures at intermediate $U$ and $f$. The intermediate-strength **E** is needed to asymmetrically morph the axisymmetric skyrmions observed at low $U$, without destroying the skyrmions by the strong electric alignment taking place at high $U$. Using intermediate frequencies avoids various electro-kinetic instabilities at low $f$ (at which skyrmions become unstable due to spatial re-distributions of ions that further alter the director field) while still allowing for out-of-equilibrium temporal evolution of **n(r)** not invariant under turning the instantaneous voltage on and off. This is because the electric field oscillation period $T_U = 1/f$ within this frequency range is comparable to the LC's rising and falling response times (within 20-100ms for our samples). Formation of clusters and chains within schools at different voltages is consistent with the nature of out-of-equilibrium elastic interactions between skyrmions at oscillating $U$ revealed by numerical modeling (Fig. 5f-m) and qualitatively discussed above by using the dipolar elastic interactions. Fine details of clusters, like inter-particle distances (Fig. 5n) vary along the $f$-axis and would be difficult to capture within a single diagram of states, but the simplified three-dimensional diagram in Fig. 7 overviews the tendencies and helps to emphasize the physical underpinnings of the observed rich out-of-equilibrium behavior, which we summarize below.

**Discussion.**

Playing a key role in skyrmion schooling, many-body elastic interactions (Figs. 4 and 5)

minimize the energetic costs of periodically-varying $\mathbf{n(r)}$ within schools by tending to position individual skyrmions such that they share the dynamic distortions and reduce the overall free energy. Motion of clusters and chains is impeded as compared to the fastest individual skyrmions (Fig. 6e,f) because of the very same sharing of asymmetric dynamic distortions between individual skyrmions, the non-reciprocal evolution of which is the source of motion. During each $T_U$=10-20ms within schooling (Figs. 4 and 5), $\mathbf{n(r)}$ never fully relaxes because the LC's 20-100ms response time is longer than $T_U$, so that the skyrmions are always asymmetric with periodically-changing preimage dipole tilts (Fig. 1j,k). When the instantaneous $U$ within $T_U$ drops to zero, the elastic torque tends to relax the $\mathbf{n(r)}$ of all skyrmions to an axially symmetric state shown in Fig. 1g, but well before this happens, competing electric and elastic torques re-morph the skyrmions back to the highly-asymmetric structures (Fig. 1j,k). While a viscous torque resists changes of $\mathbf{n(r)}$, torque balances are different in the presence of $\mathbf{E}$ and without it, making the responses to turning instantaneous $U$ on and off highly asymmetric and non-reciprocal[24]. Consequently, asymmetric skyrmions translate within a tilted director background in response to oscillating voltage, roughly antiparallel to their $\mathbf{p}_i$, much like polar granular particles translate in response to mechanical vibrations[19]. Asymmetric skyrmions synchronize $\mathbf{p}_i$ and $\mathbf{v}_i$ even before colliding by sensing each other through the long-range many-body elastic interactions.[29] Remarkably, this elasticity-enhanced synchronization can take place at packing fractions ~0.01 and inter-skyrmion distances ~10 times larger than the soliton's lateral size, consistent with the long-range nature of elastic forces. Given that collective motions can arise at carrier frequencies ~1kHz, at which ions are too slow to follow oscillating fields and their dynamics can be neglected, electro-kinetics is not a pre-requisite for the studied effects. Since numerical modeling reproduces motions of individual skyrmions[24] and their chains and clusters when using only the rotational viscosity/torque (Fig. 5f-m and Methods), it appears that flows are not essential for the collective dynamics of skyrmions, like in the "dry" types of active matter[2] (e.g. herds of cows and biological cells crawling on substrates), though such flows are locally present[23]. Backflow and electrokinetic effects enrich the collective dynamics, though their detailed study and uses are beyond the scope of our current work. Skyrmion clusters also dynamically interact to exchange and re-arrange elementary skyrmions both spontaneously and during interactions with obstacles, such as other skyrmions pinned to substrates using laser tweezers (Fig. 8 and Supplementary Movie 1). Importantly, this local bypassing of obstacles does not alter the direction of schooling, but could potentially be a useful tool in controlling collective dynamics in schools of skyrmions, as well as could potentially be extended to other active matter systems[37].

To conclude, we have demonstrated active matter formed by solitonic particle-like field configurations, with salient features of energy conversion at the individual particle level and synchronization of initially-random motion directions that leads to skyrmion schools. This schooling displays voltage-controlled self-reconfigurations of coherent motions with and without clustering. While much of the recent excitement in active matter has been generated by topological defects, which exhibit fascinating dynamics[33,34] and play key roles in living tissues[35,36], our findings demonstrate that not just singular defects, but also topological solitons can behave like active particles. Skyrmion schooling can allow for modeling diverse forms of non-equilibrium behavior, benefiting from non-biological origins and on-demand creation/elimination of skyrmions using laser tweezers[29] and providing insights into the role of topology and orientational elasticity in active matter. For example, it will be of interest to explore how giant number fluctuations arise despite of and in the presence of the self-aligning nematic fluid hosts of skyrmions with orientational elasticity and long-range inter-skyrmion interactions. Our system comprises commercially-available ingredients, with design and preparation techniques benefiting from display industry developments. It also connects the topology[29] and active matter[2] paradigms, potentially resulting in fertile new research directions at their interfaces. From a materials-applications perspective, one can envisage photonic and electrooptic materials, including displays and privacy windows, with built-in emergent responses capable of controlling light, effectively expanding potential exciting applications of more common active materials[2-22,33-36]. Being compatible with the touch-screen displays and related technologies, skyrmion schooling can be coupled to external stimuli responses and interactions with humans, potentially yielding active matter art and computer games invoking emergent behavior. Since these skyrmions can carry nanoparticle cargo[25,32], their schooling can yield active self-reconfigurable metamaterials and nanophotonic devices. Our approach can be also extended to nematic colloids[30,31], where one can potentially achieve electrically powered schooling of colloidal particles within the LC. These findings call for the development of novel active matter modeling approaches capable of handling collective behaviors of thousands-to-millions of schooling skyrmions, each with periodically-morphing complex structures of the molecular alignment field and with temporal director evolution coupled to flows and electro-kinetic effects.

## Methods

### Sample preparation

Chiral nematic LC mixtures with negative dielectric anisotropies were prepared by mixing a chiral additive (CB15, ZLI-811, or QL-76) with a room-temperature nematic host (MLC-6609 or ZLI-2806). CB15 and ZLI-2806 were purchased from EM Chemicals. ZLI-811 and MLC-6609 were purchased from Merck. The QL-76 chiral additive[38,39] was obtained from the Air Force Research Laboratory (Dayton, Ohio). Pitch, *p,* of the mixtures was controlled by varying the concentration, $c$, of the chiral additive with known helical twisting power (Table 1), $h_{HTP}$, according to the relation $p=1/(h_{HTP} \cdot c)$. Studied samples had *p=3*-10µm and *d/p* ≈ 1.[23] The LCs were mixed with ~0.1wt% of cationic surfactant Hexadecyltrimethylammonium bromide (CTAB, purchased from Sigma-Aldrich) in order to allow for inducing electrohydrodynamic instability with low-frequency applied field[32]. The CTAB doping allowed for facile generation of large numbers (thousands to millions) of skyrmions at different initial packing fractions upon relaxation of the cells from electrohydrodynamic instability (Fig. 9), though the presence of CTAB is not required for skyrmion schooling as similar dynamics could also be obtained in samples without CTAB and with skyrmions generated by laser tweezers. Furthermore, in chiral nematic cells with weak perpendicular boundary conditions and $d/p \approx 1$, skyrmions could be formed spontaneously upon quenching samples from isotropic to the LC phase and exhibited similar dynamic behavior. LC cells were constructed using glass substrates with transparent indium tin oxide conductive layers and spin-coated with polyimide coatings (SE-1211 purchased from Nissan) to impose the finite-strength perpendicular surface boundary conditions for the LC director. Spin-coating was done at 2700 rpm for 30 s. The substrates were then baked for 5 min at 90 °C and for 1 h at 190 °C to induce cross-linking of the alignment layer. The substrates were glued together, with the treated surfaces facing inward, and the cell gap was set with glass fiber segments dispersed in the ultraviolet-curable glue. The cells were cured for 60 s with an OmniCure UV lamp, Series 2000. Commercially available homeotropic cells (purchased from Instec) were also used. Electrical connections for voltage application across the depth of the LC cell were achieved by soldering leads to the ITO-electrode surfaces. Finally, the LC was heated to the isotropic phase, infiltrated into the constructed cells via capillary action and sealed with 5-minute fast-setting epoxy.

**Generation, manipulation, pinning and control of skyrmions**

Skyrmions can be formed spontaneously upon thermally quenching the sample from the isotropic to the chiral nematic phase, by relaxing the LC from electrohydrodynamic instability (Fig. 9), or by a direct one-by-one optical generation using holographic laser tweezers[23,32]. Spontaneous formation of large densities of skyrmions was achieved by inducing an electrohydrodynamic instability in cells doped with CTAB by applying low-frequency voltage of $U = 5 - 25$ V at frequencies within $2 - 10$ Hz. The control of this voltage and carrier frequency allows for the selection of the initial skyrmion packing fraction within 0.01-0.40 (Fig. 9). Laser-induced generation of individual skyrmions in all other cells without CTAB was done using optical tweezers comprised of a 1064 nm Ytterbium-doped fiber laser (YLR-10-1064, IPG Photonics) and a phase-only spatial light modulator (P512-1064, Boulder Nonlinear Systems). Using this setup, we can controllably produce arbitrary, dynamically-evolving 3D patterns of laser light intensity within the sample and generate twisted structures by means of optically-induced local reorientation of the director field known as optical Fredericks transition. Upon focusing this laser beam of power >50 mW in the midplane of the cell, the local LC director realigns away from the far-field background by coupling to the optical-frequency electric field of the laser beam. Skyrmions were individually generated by laser tweezers at ~50 mW power and selectively pinned to the substrate surface to act as obstacles in desired locations using powers of 70-150 mW. The LC's tendency to twist makes skyrmions energetically favorable at given confinement conditions, so that they form spontaneously after the uniform background of homeotropic cells is distorted by electrohydrodynamic instabilities or laser-induced realignment (after voltage or laser light are turned off, as shown in Fig. 9), whereas control of distortions in these two cases allows for defining initial densities and locations of skyrmions. In order to morph skyrmions and power their motions via macroscopically-supplied energy, electric field was applied across the cell using a homemade MATLAB-based voltage-application program coupled with a data-acquisition board (NIDAQ-6363, National Instruments)[24].

The means of controlling dynamics of skyrmions include voltage driving schemes, selection of LCs with different material parameters and chiral additives (Table 1), design of LC cell geometry, and strength of surface boundary conditions, etc. Collective motion effects could be obtained both when simply using low frequency oscillating field (yielding field oscillations at time scales comparable to the LC's response time) and when modulating high-frequency carrier signals (e.g. at 1kHz) at modulation frequencies that again yield modulation periods comparable to the LC's response time. The large range of possibilities exists in terms of controlling the skyrmion schooling by varying electric signal waveforms, modulation and carrier frequencies and so on, but detailed exploration of all these possibilities is outside the scope of our present study.

**Numerical modeling**

We computer-simulated structures of skyrmions and their self-assemblies at experimental conditions (Figs. 1 and 5) by using LC free energy with elastic and electric coupling terms[24,40-41]:

$$W = \int \left\{ \begin{matrix} \frac{K_{11}}{2} (\nabla \cdot \boldsymbol{n})^2 + \frac{K_{22}}{2} [\boldsymbol{n} \cdot (\nabla \times \boldsymbol{n}) + q_0]^2 + \frac{K_{33}}{2} [\boldsymbol{n} \times (\nabla \times \boldsymbol{n})]^2 \\ -K_{24} \{ \nabla \cdot [\boldsymbol{n}(\nabla \cdot \boldsymbol{n}) + \boldsymbol{n} \times (\nabla \times \boldsymbol{n})] \} - \frac{\varepsilon_0 \Delta \varepsilon}{2} (\mathbf{E} \cdot \boldsymbol{n})^2 \end{matrix} \right\} dV \tag{1}$$

where Frank elastic constants $K_{11}$, $K_{22}$, $K_{33}$ and $K_{24}$ represent the elastic costs for splay, twist, bend and saddle splay deformations of $\mathbf{n(r)}$, respectively. The chiral wavenumber of the ground-state chiral nematic mixture is defined as $q_0 = 2\pi/p$ and $\Delta\varepsilon$ is the dielectric anisotropy. We take $K_{24} = K_{22}$, as in previous studies[32], whereas all other material parameters used correspond to the experimental values (Table 1). As the applied voltage is modulated, the competing electric and elastic torques are balanced by a viscous torque associated with rotational viscosity, $\gamma$, that opposes the fast rotation of the director[1]. The resulting director dynamics is governed by a torque balance equation[1], $\gamma \partial n_i / \partial t = -\delta W / \delta n_i$, from which both the equilibrium $\mathbf{n(r)}$ and the effective temporal evolution of the director field towards equilibrium are obtained for the director, $n_i(t)$ where $n_i$ is the component of $\mathbf{n}$ along the $i^{th}$ axis ($i$ = x, y, z). As in experiments, we start our computer simulations from skyrmions embedded in a homeotropic LC background, for which the equilibrium director structure is obtained by minimizing free energy in Eq. (1) at no external fields (Fig. 1g), as detailed in our previous studies.[24,29] Then, we minimize free energy to obtain skyrmion's field configuration's at various applied voltages (Fig. 1j,k) when starting from the skyrmion structure at no fields (Fig. 1g) within the computational volume as the initial condition. In a similar way, to obtain multi-skyrmion clusters or chains, we start from minimizing free energy at $U=0$ for seven (Fig. 5f) or five (Fig. 5j) axisymmetric skyrmions embedded in the homeotropic background of the computational volume and then use these structures as initial conditions to obtain clusters in the corresponding applied fields. Similar to experiments, skyrmions self-organize into clusters and chains (Fig. 5f-m) at corresponding voltages and translate laterally by about $p/2$ each time as we effectively turn voltage on and off and minimize free energy at the corresponding conditions. Periodic turning voltage on and off that corresponds to $T_U$ results in a periodic nonreciprocal director-field evolution that yields an asymmetric shifting of the skyrmions between the voltage-on and voltage-off states, resulting in a displacement within the computational volume[24], similar to that seen in experiments. Such periodic displacements add to yield lateral translations of both individual skyrmions and their clusters and chains (Fig. 5f-m). Similar to experiments, the velocity vectors that we obtain from analyzing displacements of skyrmionic $\mathbf{n}(\mathbf{r})$-structures are anti-parallel to the preimage vectors $\mathbf{p}_i$. Dynamic evolution of $\mathbf{n}(\mathbf{r})$ has been used to derive the intermediate states between the voltage-on and voltage-off states by taking snapshots of the director field during the process of evolution towards equilibrium within each voltage modulation period $T_U$[24]. Once the director structures are obtained, we utilize a Jones-matrix method[23,24] to generate polarizing optical micrographs for experimental parameters such as sample thickness, optical refractive index anisotropy, and $p$ (Table 1).

**Optical microscopy, video characterization and data analysis**

Images and videos were obtained using charge-coupled device cameras Grasshopper (purchased from Point Grey Research, Inc.) or SPOT 14.2 Color Mosaic (purchased from Diagnostic

Instruments, Inc.), which were mounted on an upright BX-51 Olympus microscope. Dry 2x, 4x, 10x, and 20x objectives (with numerical apertures ranging from 0.3 to 0.9) were used, with different relative orientations of polarizers adjusted to increase contrast between the skyrmionic structures and the background. This contrast was integral to successfully analyze the polarizing micrographs for skyrmion motion using the open-source ImageJ/FIJI software (obtained from the National Institute of Health). Built-in particle-tracking tools were applied, through which skyrmion positional information and skyrmion number density were extracted for each frame. Then, the data analysis and plotting in MATLAB software (obtained from MathWorks) were performed to characterize trajectory pathways, velocity and polar order parameters, giant-number fluctuation scaling, and density fluctuations. The temporal evolution of both polar and velocity order was characterized by analyzing the positional data for skyrmion motion between frames of the videos. The velocity vector for an individual skyrmion $\mathbf{v}_i$ was defined by drawing a vector between the skyrmion's positions in consecutive frames of the video, pointing along the direction of motion[4]. The polar preimage vector for each skyrmion $\mathbf{p}_i$ was defined by drawing a normalized vector between the south-pole and north-pole preimages. The giant-number fluctuations and scaling trends were analyzed using the skyrmion number density data obtained by means of the ImageJ/FIJI particle-counting features[15]. This was done by doing skyrmion number density analysis with time for fifteen areas of different sizes, ranging from 25µm x 25µm to 1250µm x 1250µm, for each experimental video. Figure 4f represents the number density fluctuation analysis for one representative 400µm x 400µm region as an example. Various regions within the samples were probed for each experimental video and a composite of 3-5 videos were analyzed for each case (individual, clusters, and chains of skyrmions), resulting in ~120 data points each to represent the different sample areas. The time period over which the fluctuations were characterized for each video was within 150-180 s. The density data points by area were compiled and plotted as log-log plots of the mean particles $<N>$ and root mean square $\Delta N = <(N-<N>)^2>^{1/2}$ in Figures 4e and 5e.

**Data availability** All datasets generated and analyzed during the current study are available from the corresponding author on reasonable request.

**Code availability** Matlab codes generated and analyzed during the current study are available from the corresponding author on request.

## Acknowledgments

This research was supported by the National Science Foundation through grants DMR-1810513 (research), DGE-1144083 (Graduate Research Fellowship to H.R.O.S.) and ACI-1532235 and ACI-1532236 (RMACC Summit supercomputer used for the numerical modeling). We thank P.

Ackerman, M. Bowick, M. Cates, A. Hess, T. Lubensky, D. Marenduzzo, S. Ramaswamy, M. Ravnik, M. Tasinkevych, J. Toner, J. Yeomans and S. Zumer for discussions and P. Ackerman for technical assistance. We thank Corbin Sohn for the assistance with taking the photograph of a school of fish shown in Fig.1a, which is part of the Sohn family collection.

**Author Contributions** H.R.O.S. performed experiments (with assistance from C.D.L.) and numerical modeling. H.R.O.S., C.D.L. and I.I.S. analyzed data. I.I.S. wrote the manuscript (with the input from all authors), conceived the project, designed experiments and provided funding.

**Competing interests** Authors declare no competing interests.

**Correspondence and requests for materials** should be addressed to I.I.S.

**Figures**

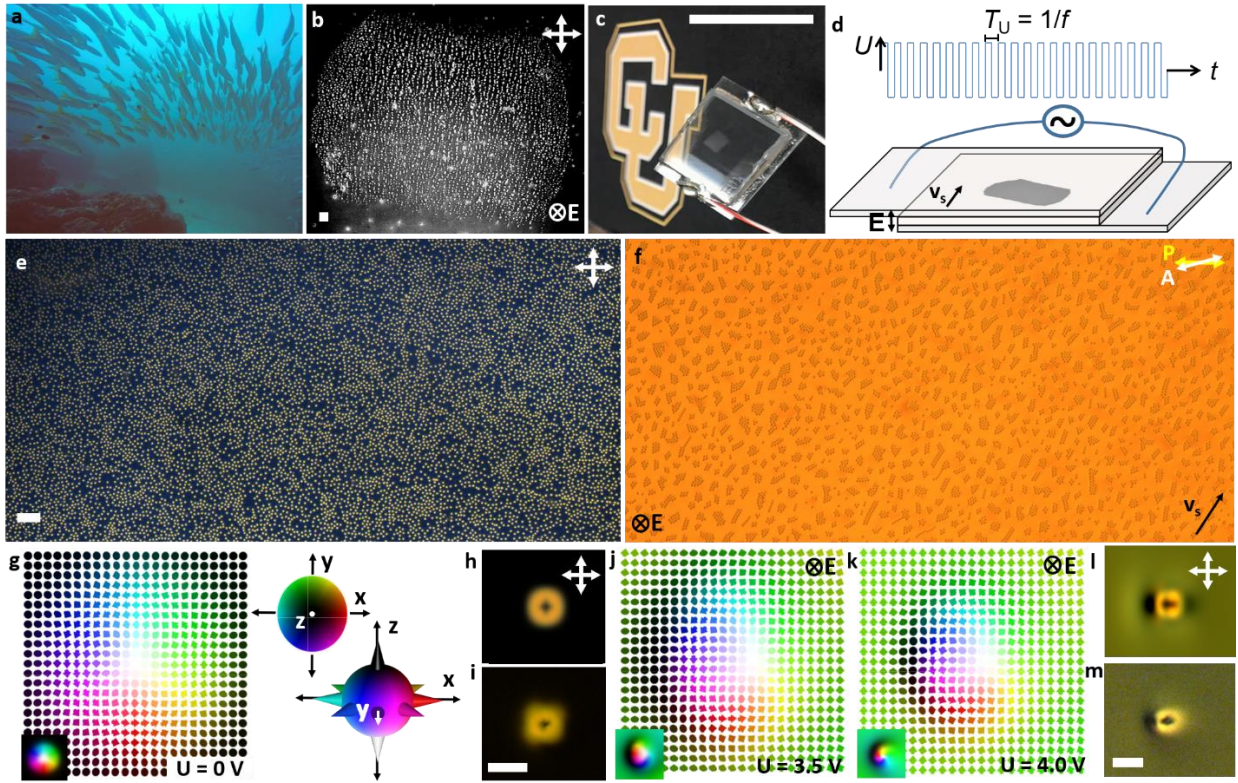

**Figure 1 | Skyrmions and schooling. a,** Underwater photograph of fish schooling. **b,** Zoomed-out grayscale polarizing optical micrograph of a school of skyrmions. The scale bar in the bottom left is 100 μm. **c,** Photograph of the skyrmion school shown in (**b**), visible with the unaided eye as the cloudy region within the sample. The scale bar is 1 inch. **d,** Schematic of a sample with voltage application across the LC by using transparent electrodes on the inner surfaces of the confining substrates. The dark region within the schematic represents the moving skyrmion school shown in (**b,c**). **e, f,** Low-magnification experimental polarizing micrographs of a region within the school at $U$=0V (**e**) and $U$=3.5V (**f**). The scale bar in the bottom left of (**e**) is 100 μm. **g,** Numerically-simulated $\mathbf{n(r)}$ of an individual skyrmion at $U$=0 shown by arrows colored according to points on the two-sphere $\mathbb{S}^2$, as shown in the right-side inset. **h,i,** Computer-simulated polarizing micrograph (**h**) closely matches its experimental counterpart (**i**). Numerically-simulated $\mathbf{n(r)}$ of skyrmions at $U$=3.5V (**j**) and $U$=4.0V (**k**), with corresponding computer-simulated (**l**) and experimental (**m**) polarizing micrographs at $U$=4.0V. Bottom left insets in (**g, j,k**) are smoothly-colored representations of $\mathbf{n(r)}$. The scale bars in (**i**) and (**m**) are 10 μm . Directions of $\mathbf{E}$, skyrmion motion velocity direction $\mathbf{v_s}$, $U$, $T_U$, double arrows denoting polarizer orientations and time $t$ are marked throughout. Numerical simulations are based on material parameters of LC mixture of nematic MLC-6609 and right-handed chiral additive CB-15 (Methods) and $d \approx p \approx 10$ μm.

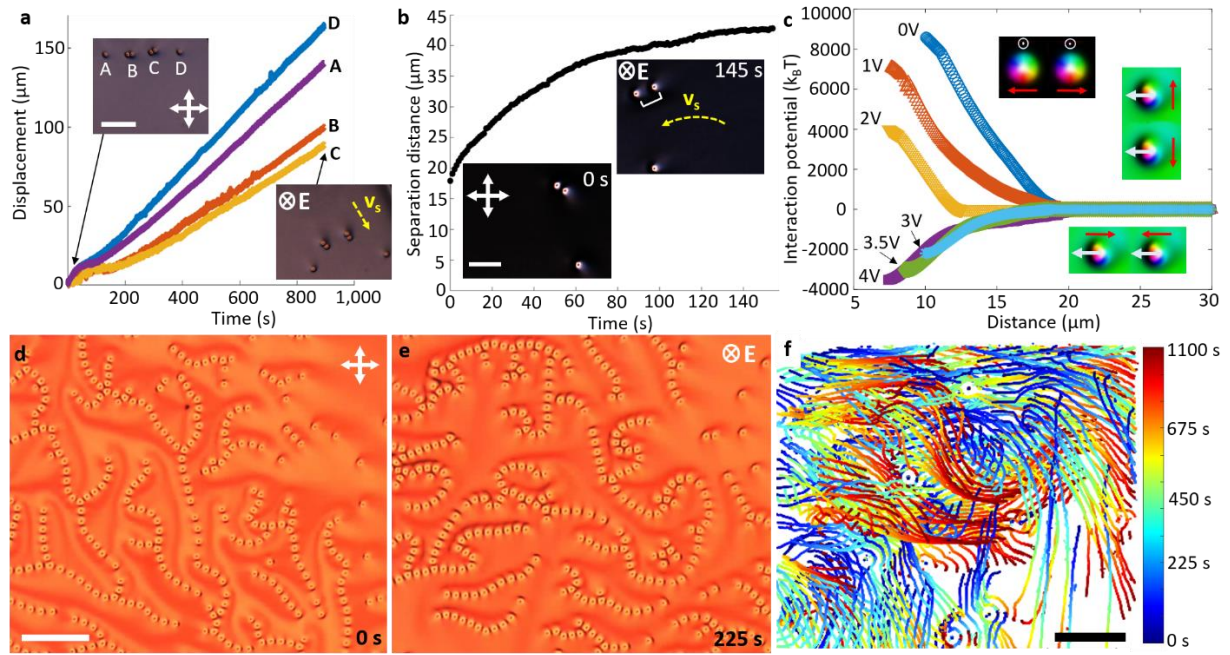

**Figure 2 / Pair interactions and motions of individual and two-to-hundreds assemblies of skyrmions. a**, Comparison of motions of individual and pairs of skyrmions labeled as in the video frames shown in insets. Applied voltage is $U$=4.5V, oscillated at $f$=2Hz; high-frequency 1 kHz carrier signal is used (see Methods), indicating that these motions are not caused solely by the dynamics of ions[24]. The scale bar is 100 μm . **b**, Pair separation distance versus time during skyrmion motion along the arched trajectoris, with insets showing skyrmions in the first and last frames and the pair separation distance labeled on the last frame. Motion was powered by $U$=4.5 V and at $f$=1 Hz oscillation frequency, with high carrier frequency of 1 kHz. Velocity directions $\mathbf{v}_s$ are marked in insets of (**a**,**b**) with yellow dashed arrows. The scale bar is 50 μm. **c**, Pair interaction potential for skyrmions in motion at various voltages and at $f$=75Hz. Insets show smoothly-colored simulated $\mathbf{n}(\mathbf{r})$ configurations that follow the color scheme in Fig. 1g and reveal the presence of attractions or repulsions. Gray arrows mark the preimage vectors and red arrows represent attractive or repulsive forces between skyrmions. **d**, **e**, Frames from video showing temporal evolution of chains with hundreds of skyrmions at $U$=4.2V oscillated at $f$=2 Hz. **f**, Time-colored trajectories representing skyrmion motions over a time period of 1100 s, with the right-side inset showing the time-color scheme. The scale bars in (**d**) and (**f**) are 100 μm. The chiral LC is MLC-6609 doped with left-handed chiral additives (**a**,**b**) ZLI-811 and (**d**, **e**) QL-76. White double arrows denote crossed polarizers for all micrographs in this figure. **E** is marked in (**e**) and in the last frame of insets in (**a**,**b**). The elapsed time is marked in the bottom-right of frames.

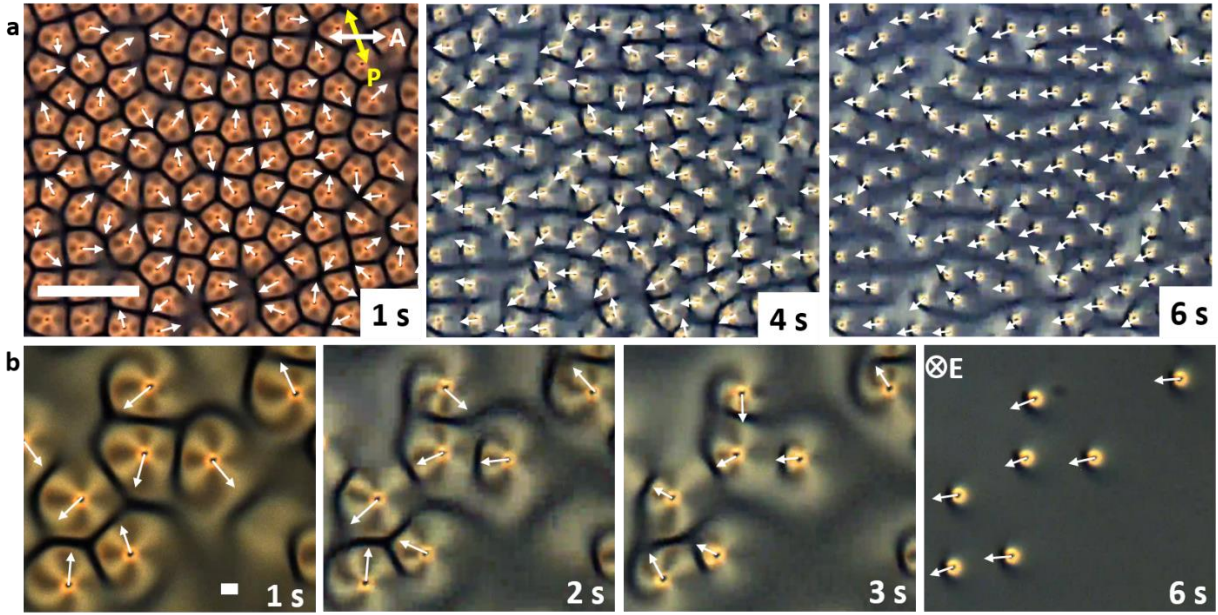

**Figure 3 | Temporal evolution of skyrmion velocity vector orientations.** This analysis is done at the onset of the ordering transition corresponding to Fig. 4d of the main text. **a,b,** Polarizing micrographs of (**a**) a higher number density region and (**b**) a lower number density region within the same sample at elapsed times marked on the images. White arrows point along the directions of skyrmion motion, representing the velocity unit vectors, $\mathbf{v}_i$. The applied voltage is $U$=3.5 V. The scale bars are (**a**) 100 μm and (**b**) 10 μm. Double arrows denote uncrossed polarizer and analyzer orientations and direction of voltage application in marked in the last frame of (**b**). The chiral LC is the nematic host ZLI-2806 doped with CB-15.

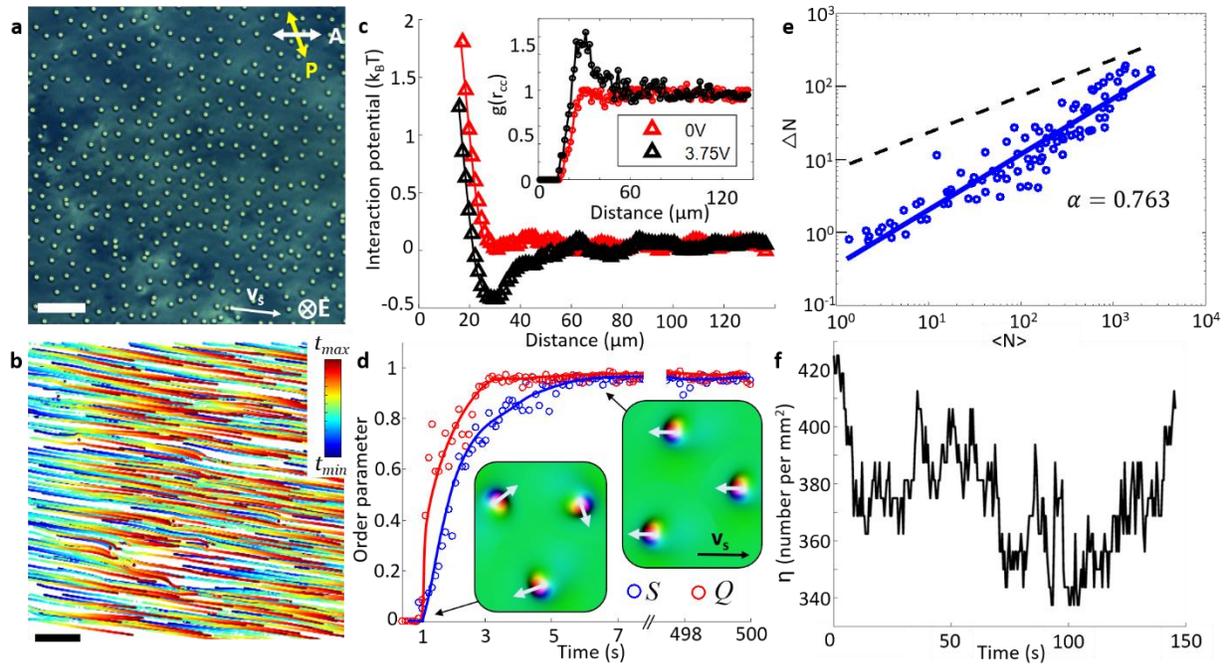

**Figure 4 | Coherent skyrmion motion in a school. a**, Experimental polarizing micrograph of moving skyrmions in a school without clustering for initial packing fraction of 0.09 by area at $U$=3.75 V and $f$=50 Hz. Polarizer and analyzer orientations are marked with double arrows and directions of **E** and skyrmion motion are marked in the bottom-right corner. **b**, Trajectories of skyrmion motions color-coded with time according to the scheme shown in the right-side inset (where $t_{min}$=0 and $t_{max}$=72 s); these trajectories correspond to Supplementary Movie 5. The scale bars in (**a**) and (**b**) are 100 μm**. c**, Interaction potential (extracted from the radial distribution function g($r_{cc}$) shown in the inset) of center-to-center distances $r_{cc}$ for schooling skyrmions at conditions like in part (**a**). **d**, Evolution of $S$ and $Q$ with time. Insets schematically show corresponding configurations colored according to the scheme in the inset of Fig.1g. **e**, Giant number fluctuation analysis using a log-log plot of $\Delta N$ versus $<N>$; black dashed line indicates a slope of 0.5 for reference. **f**, An example of number density fluctuation during motion for a 400μm x 400μm sample area. Such schools of individually dispersed, collectively moving skyrmions are observed at $U$=2.5-4.25 V and initial packing fractions within 0.01-0.09 by area. The LC is ZLI-2806 doped with right-handed additive CB-15 (Methods).

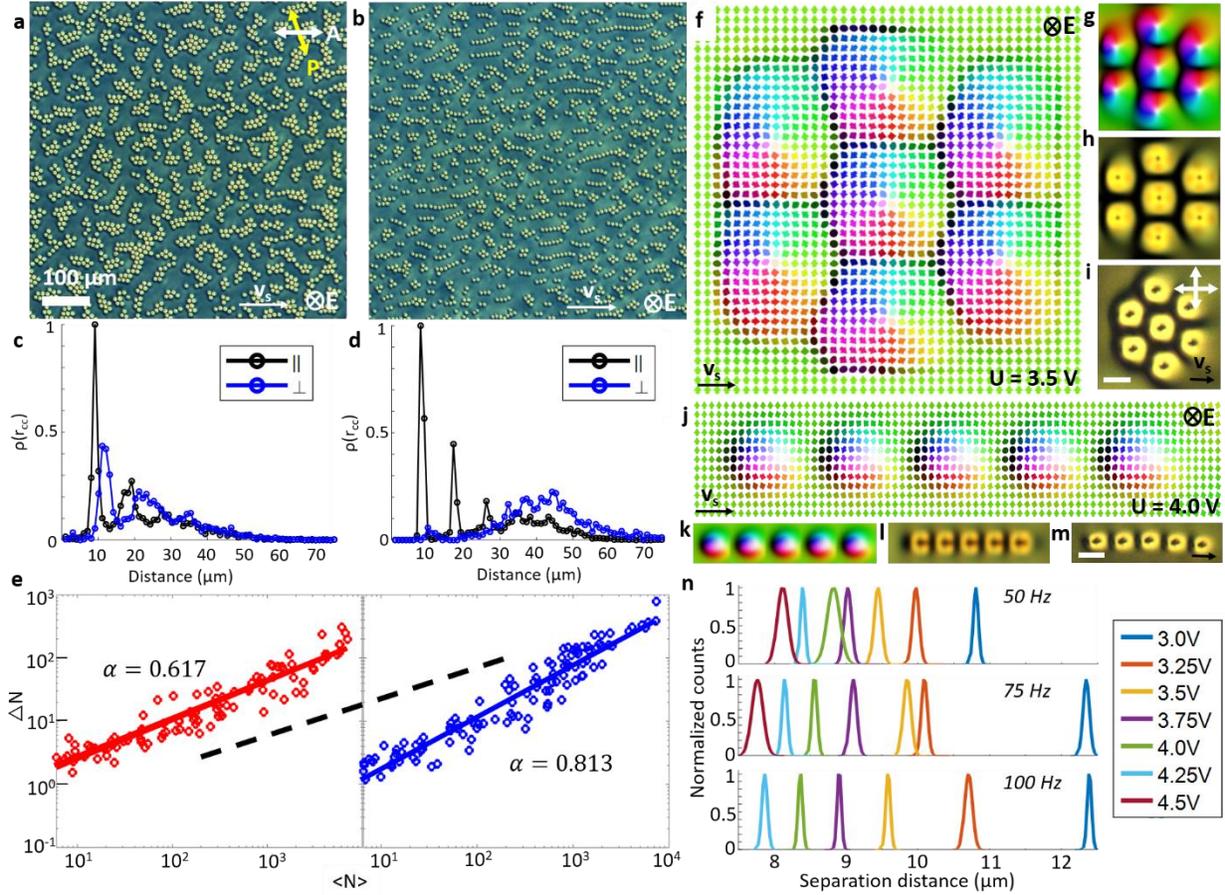

**Figure 5 | Schools of skyrmions with internal clusters and chains. a,b,** Polarizing micrographs show schooling of skyrmions with clusters at $U$=3.5V (**a**) and chains at $U$=4.0V (**b**), where $f$=50Hz. The scale bar in (**a**) is 100 µm. **c,d,** Normalized skyrmion count distributions $\rho_{\perp}(r_{cc})$ and $\rho_{\parallel}(r_{cc})$ for directions perpendicular and parallel to the motion directions for clusters (**c**) and chains (**d**) shown in (**a**) and (**b**), respectively. **e,** Log-log plots of $\Delta N$ versus $<N>$ for clusters (red) and chains (blue); linear fits with slope $\alpha$ values and a black dashed line with $\alpha = 0.5$ (for reference) are displayed. **f,g,** Computer-simulated moving cluster of seven quasi-hexagonally-assembled skyrmions at $U$=3.5V displayed using arrows colored according to the color scheme in Fig.1g (**f**) and corresponding smoothly-colored representation of **n(r)** (**g**). **h,** Simulated and **i,** experimental polarizing optical micrographs of a similar moving cluster. **j,k,** Simulated moving linear chain of five skyrmions at $U$=4.0V (**j**) with corresponding smoothly-colored **n(r)** representation (**k**). **l,** Simulated and **m,** experimental polarizing micrographs of the moving chain. Scale bars in (**i**) and (**m**) are 10 µm. Upon turning voltage off and on again while minimizing **n(r)**, the clusters in (**f,j**) shift laterally within the computational volume by about ~0.5$p$. **n,** Experimental normalized distributions for inter-skyrmion distance in chains during motion versus $U$ for $f$=50, 75 and 100 Hz. Polarizer orientations (double arrows), **v$_s$** and **E** are labeled throughout. The LC is the nematic ZLI-2806 doped with the chiral additive CB-15.

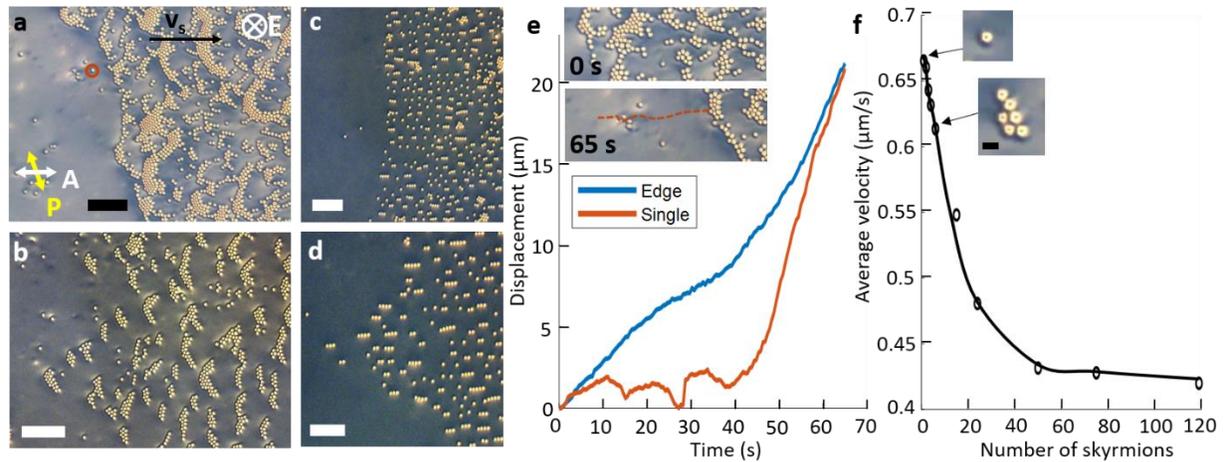

**Figure 6 | Edges of dynamic skyrmion schools. a**, Polarizing micrograph, extracted as a still frame from Supplementary Movie 6, displaying the edge of a school of skyrmions. **b-d**, Polarizing micrographs displaying different well-defined school boundaries. The scale bars in (**a-d**) are 100 μm. **e**, Displacement of a single skyrmion compared to the school's edge, derived from frames of the Supplementary Movie 6, with insets showing the moving edge with time. The trajectory of a single skyrmion (highlighted with an orange circle) in (**a**) as it catches up to the edge is shown in the lower inset with an orange dotted line. **f**, Velocity of the skyrmion clusters versus the number of constituent skyrmions within the schools. The directions of $\mathbf{v_s}$, uncrossed polarizer and analyzer (double arrows) and **E** are the same throughout as marked in (**a**). The motion takes place at $f$=50Hz and (**a, c**) $U$=3.5V and (**b, d**) $U$=4.0V. The chiral LC mixture is ZLI-2806 doped with CB-15 (Table 1).

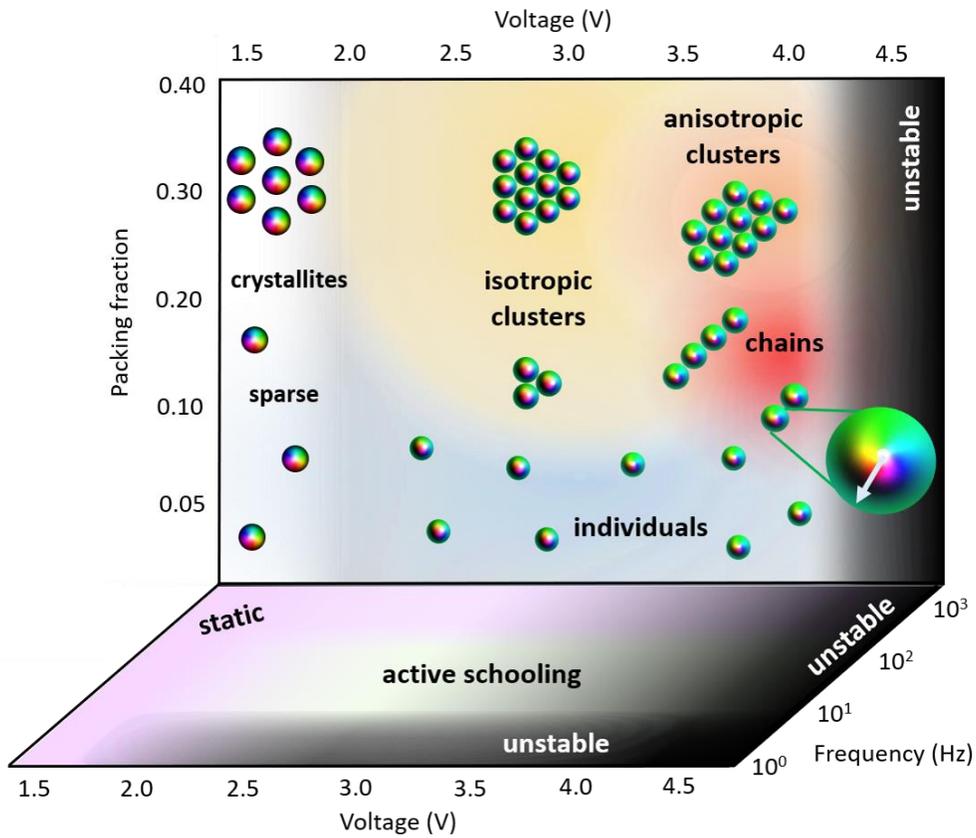

**Figure 7 / Diagram of static and dynamic skyrmion assemblies and schools.** A diagram of experimentally self-assembled dynamic and static structures versus packing fraction, $f$ and $U$. The configurations shown are consistent across all $f$ at which skyrmions are stable and shown as a single vertical panel for simplicity. Schematics of skyrmion assemblies in the insets are colored according to the scheme in the inset of Fig.1g, with the enlarged inset displaying the preimage vector of a skyrmion at $U$=4.0V.

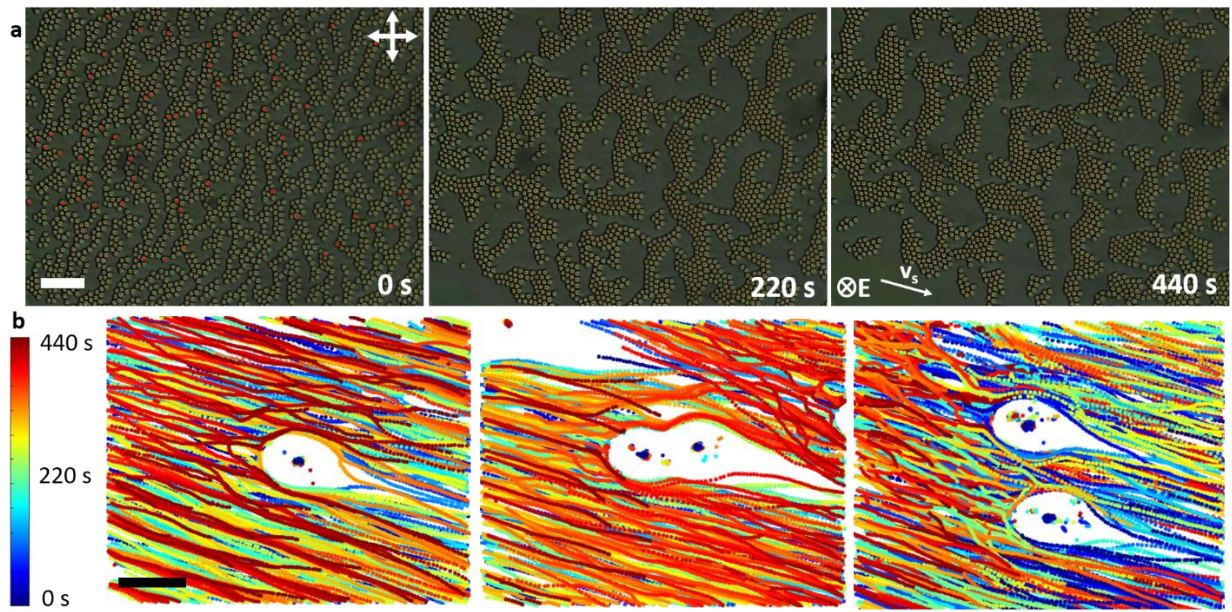

**Figure 8 | Collective motion of skyrmion clusters within schools while bypassing obstacles. a**, Polarizing micrographs of clustered skyrmion motion taken from the Supplementary Movie 1, where some skyrmions are pinned to the substrate and act as stationary obstacles (marked with red points in the first frame). The motion is powered by $U$=3 V at 60 Hz. Polarizer orientations are marked with white double arrows; direction of motion and electric field are marked on the last frame of (**a**). The elapsed time is shown in the bottom-right corners of micrographs. The scale bar is 100 µm. The chiral LC is the nematic host ZLI-2806 doped with CB-15. **b**, Examples of trajectories of skyrmion motion avoiding pinned obstacles, color-coded by the elapsed time according to the scheme shown on the left-side inset. Time-coded trajectories reveal details of skyrmions bypassing one obstacle (left), two connected obstacles (middle) and two separated obstacles (right). The scale bar is 50 µm.

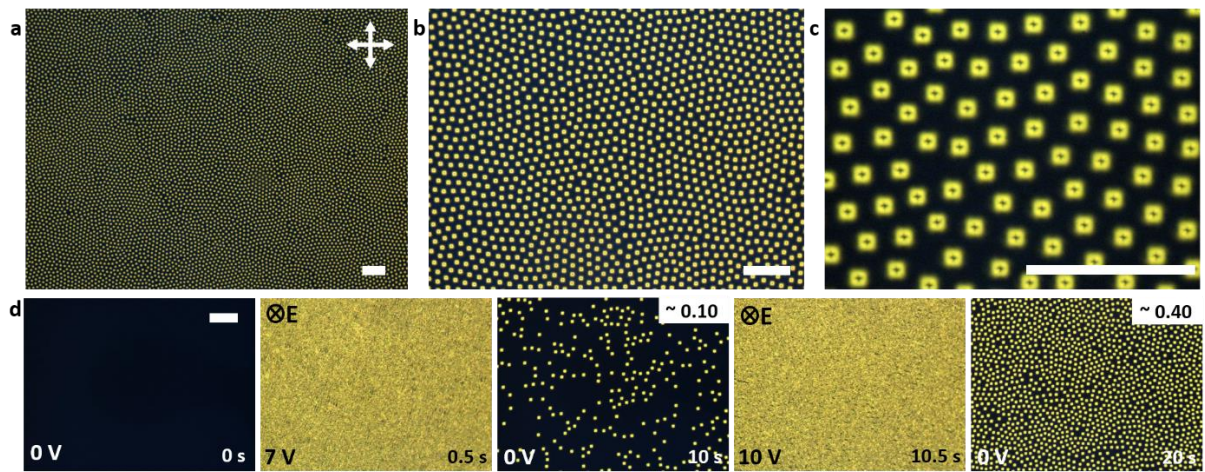

**Figure 9 | Generation of skyrmions by relaxing electrohydrodynamic instabilities. a-c,** Polarizing micrographs (at different magnifications) of a dense skyrmion array created by relaxing the electrohydrodynamic instability. The scale bars are 100 μm. **d,** Polarizing micrographs showing the as-prepared sample initially without skyrmions at no applied fields (left image), electrohydrodynamic instability at $U$=7V (2nd image from the left), skyrmions at low number density forming upon switching back to $U$=0 and relaxing the sample from electrohydrodynamic instability (3rd image from the left), electrohydrodynamic instability in the same sample area upon again applying $U$=10V (4th image from the left), and, finally, skyrmions at high number density forming upon switching back to $U$=0 again and allowing the instability to relax (5th image from the left). These images illustrate the control of the initial skyrmion packing fraction (by area, displayed in the top-right of the images), where the amplitude of $U$ used to generate electrohydrodynamic instabilities correlates with the ensuing skyrmion number densities. White double arrows denote crossed polarizer orientations. Elapsed time, skyrmion packing fraction by area, applied voltage, and direction of **E** are marked in the corners of micrographs. The used electric field frequency was 10Hz.

| Material/Property | MLC-6609 | ZLI-2806 |
|---|---|---|
| $\Delta\varepsilon$ | -3.7 | -4.8 |
| $h_{HTP}$ of CB-15 ($\mu m^{-1}$) | | +5.9 |
| $h_{HTP}$ of ZLI-811 ($\mu m^{-1}$) | -10.5 | -8.3 |
| $h_{HTP}$ of QL-76 ($\mu m^{-1}$) | -60 | |
| $K_{11}$ (pN) | 17.2 | 14.9 |
| $K_{22}$ (pN) | 7.5 | 7.9 |
| $K_{33}$ (pN) | 17.9 | 15.4 |
| $\gamma$ (mPas) | 162 | 240 |
| $n_{ext}$ | 1.5514 | 1.518 |
| $n_{ord}$ | 1.4737 | 1.474 |
| $\Delta n$ | 0.0777 | 0.044 |

**Table 1**. **Material properties of nematic hosts and chiral additives**. Parameters reported include dielectric anisotropy, $\Delta\varepsilon$, elastic $K_{11}, K_{22}, K_{33}$ constants, rotational viscosity, $\gamma$, extraordinary ($n_{ext}$) and ordinary ($n_{ord}$) refractive indices, optical anisotropy, $\Delta n$, and helical twisting powers for each chiral additive in the corresponding nematic hosts, $h_{HTP}$. Positive values of $h_{HTP}$ correspond to right-handed chiral additives and negative values of $h_{HTP}$ correspond to left-handed chiral additives.

# Supplementary Information

## Supplementary Figures

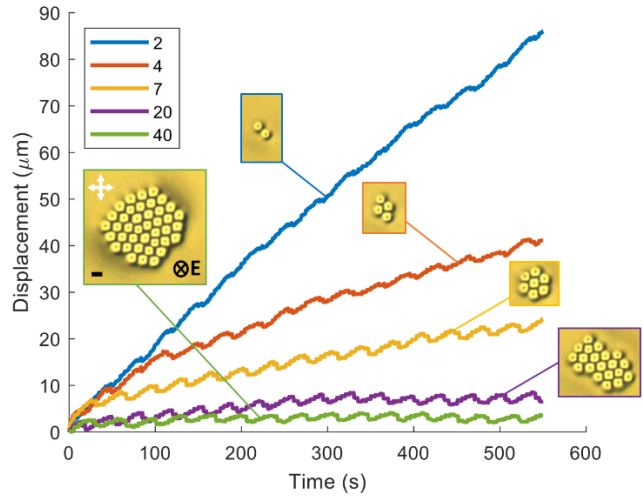

**Supplementary Figure 1 | Skyrmion displacement versus time for different clusters.** Displacement analysis of skyrmion clusters versus time at $U$=3.75 V at 2Hz oscillation frequency and at carrier frequency of 1kHz. The legend indicates the number of skyrmions in each cluster. Crossed polarizer orientations for all insets are marked with double arrows and the direction of **E** is shown. The scale bar is 10 µm. The LC is MLC-6609 doped with ZLI-811.

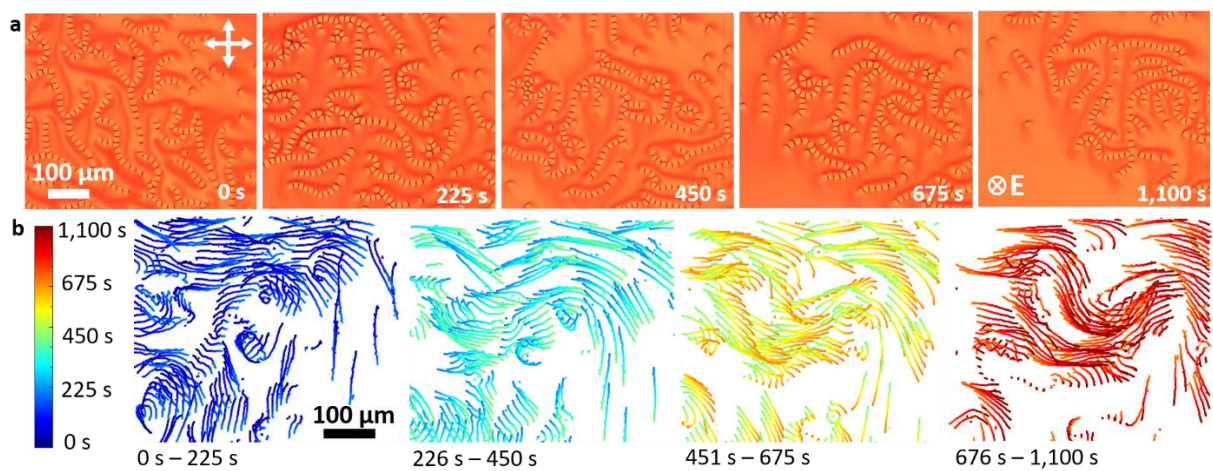

**Supplementary Figure 2 | Details of the chain-like swirling motion of skyrmions**. **a**, Temporal evolution of skyrmion chains at packing fraction ~0.005 and $U$=4.2V oscillated at $f$=2 Hz. **b**, Time-colored trajectories representing skyrmion motions between the frames displayed in (**a**), with the left-side inset showing the time-color scheme for corresponding time periods marked beneath the plots. White double arrows denote crossed polarizers for all micrographs in this figure. The direction of **E** marked in the last frame of (**a**) corresponds to all figure parts. The elapsed time is marked on the frames. The scale bars are 100 μm. The chiral LC is MLC-6609 doped with QL-76.

# Supplementary Movies

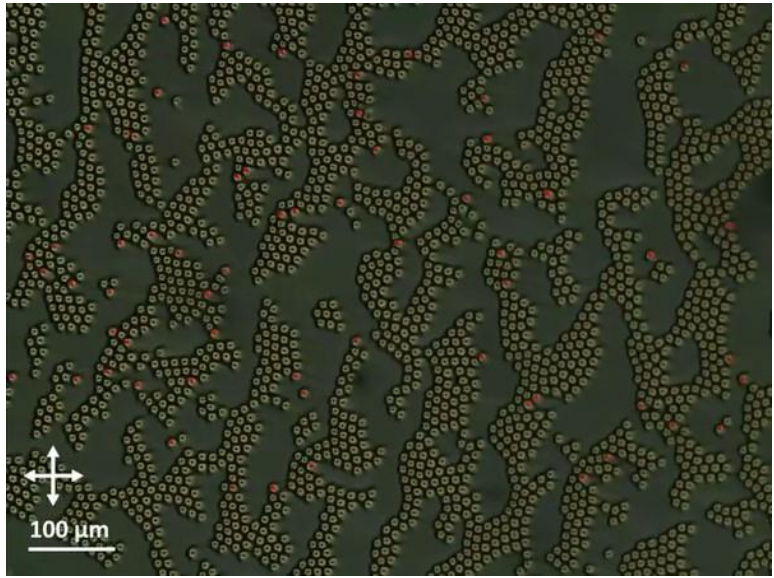

**Supplementary Movie 1 |** Collective motion of dense clusters of skyrmions moving and rearranging together to avoid randomly-pinned obstacles (stationary skyrmions, marked by red dots). The motion is powered at $U$=3V and $f$=60 Hz. The video is played at 20x speed and the actual elapsed time is 405 s. White double arrows mark the crossed polarizer orientations. The chiral LC mixture is the nematic host ZLI-2806 doped with CB-15.

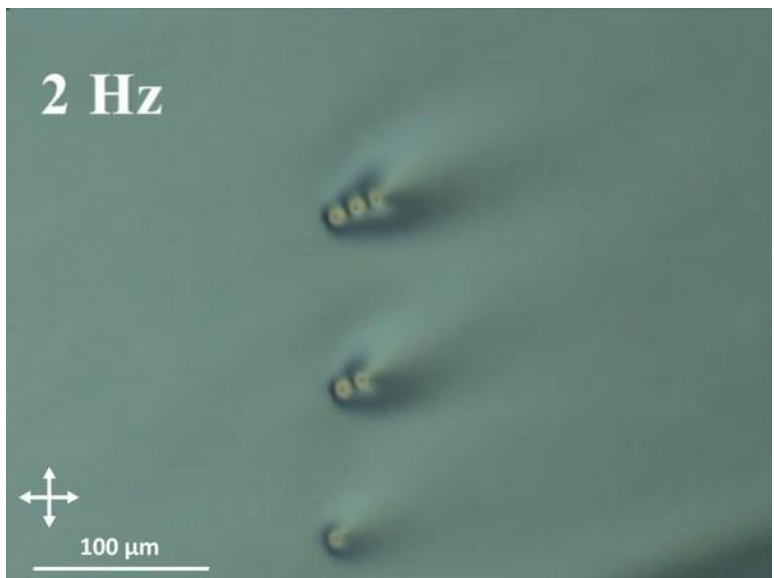

**Supplementary Movie 2 |** Skyrmion chain motion race between a trio-chain, a pair, and an individual skyrmion. The motion is powered by $U$=4.5 V voltage with oscillation at 2 or 8 Hz, as noted in top left of video frames, with carrier frequency of 1 kHz. The video is sped up 50 times. The net elapsed time is 29 minutes and 46 s. Crossed polarizer orientations are marked with white double arrows. The chiral LC mixture is the nematic MLC-6609 doped with the chiral additive ZLI-811.

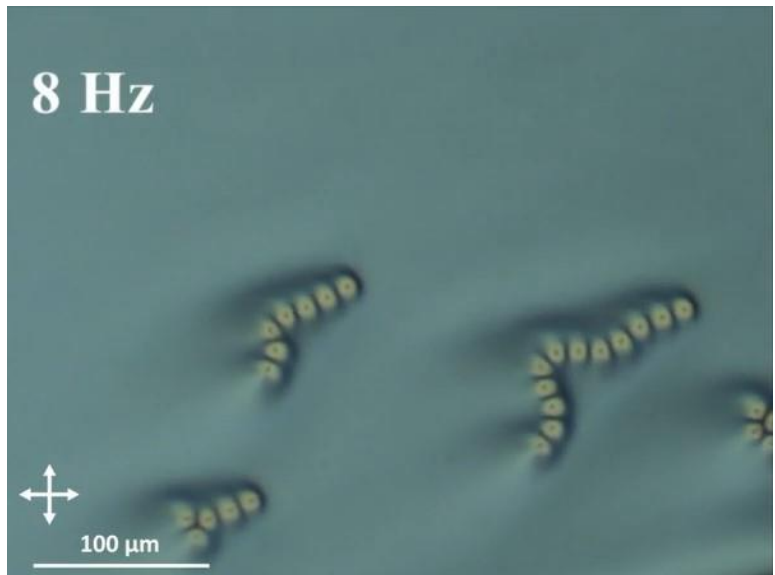

**Supplementary Movie 3 |** Kinetic out-of-equilibrium self-assembly and re-assembly of skyrmion superstructures. The motion is powered at a $U$=4.5 V voltage, with oscillation at 2 or 8 Hz, as noted in top left of video frames; the carrier frequency is 1 kHz. The video is sped up approximately 100 times and the total elapsed time is 1 hour, 10 minutes and 18 s. White double arrows denote polarizer orientations. The LC is the nematic host MLC-6609 doped with the chiral additive ZLI-811.

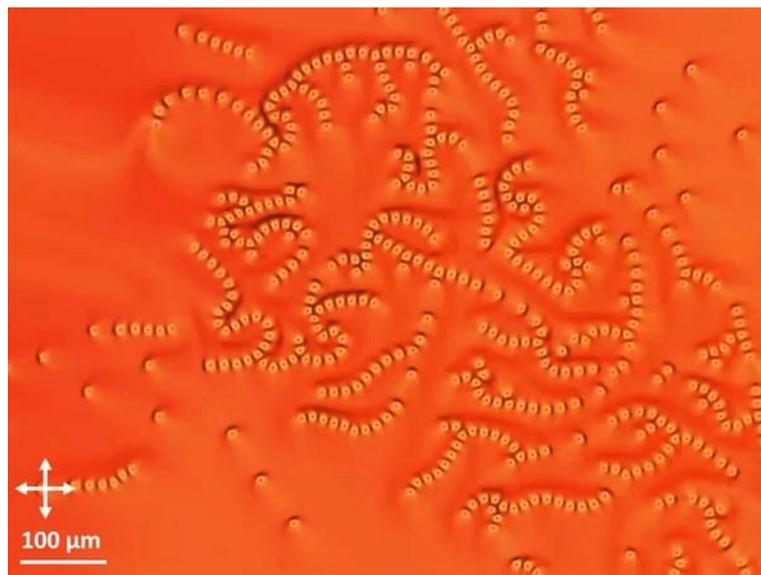

**Supplementary Movie 4 |** Temporal evolution of self-assembled meandering chain motion in a swirling pattern shown in Fig.2d-f and Supplementary Figure 2. The motion is powered with $U$=4.2V at $f$=2 Hz oscillation, with 1 kHz carrier frequency. The video is sped up 50 times the actual elapsed time is 17 minutes. White double arrows represent the crossed polarizer orientations. The chiral LC is MLC-6609 doped with the chiral additive QL-76.

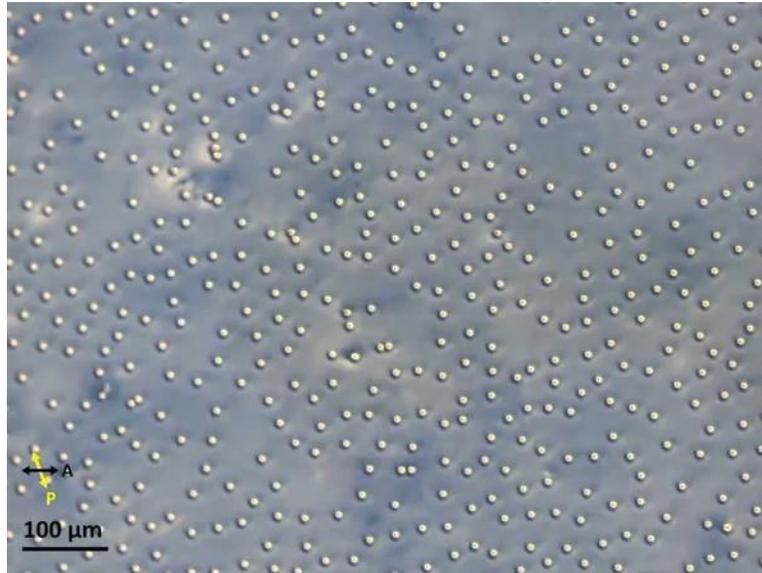

**Supplementary Movie 5 |** Large-scale collective motion of individually dispersed skyrmions within a school, powered at $U$=3.75 V and $f$=50 Hz. The video clip is sped up 3 times. The elapsed time is 65 s. Polarizer and analyzer orientation are marked with white and yellow double arrows. The chiral mixture is the nematic host ZLI-2806 doped with the chiral additive CB-15.

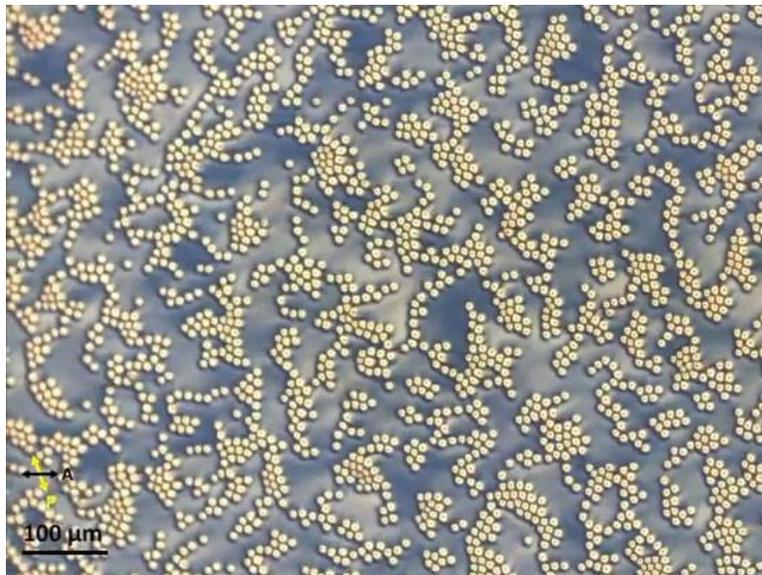

**Supplementary Movie 6 |** Large-scale collective motion of dynamically self-assembled skyrmion clusters powered with $U$=3.5 V at $f$=50 Hz. The video is sped up 10 times. The elapsed time is 234 s. Polarizer and analyzer orientations are marked with white and yellow double arrows. The chiral mixture is the nematic host ZLI-2806 doped with the chiral additive CB-15.

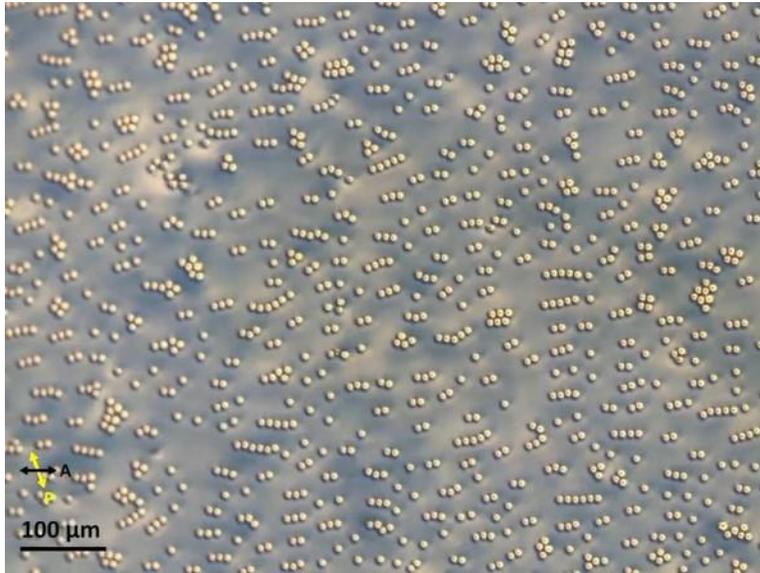

**Supplementary Movie 7 |** Large-scale collective motion of self-assembled chains of skyrmions electrically powered at $U$=4.0 V and $f$=50 Hz. The video is sped up 3 times. The actual elapsed time is 51 s. Polarizer and analyzer orientations are marked with white and yellow double arrows. The chiral mixture is the nematic host ZLI-2806 doped with the chiral additive CB-15.